# Haptics in Computer Music: a Paradigm Shift

Nicolas Castagne, Claude Cadoz, Jean-Loup Florens, Annie Luciani

Computer Arts Laboratory ACROE-ICA
Institut National Polytechnique de Grenoble,
38 000 Grenoble, France
{castagne, cadoz, florens, luciani}@imag.fr

**Abstract.** With an historical point of view combined with a bibliographic overview, the article discusses the idea that haptic force-feedback transducers correspond with a paradigm shift in our real-time digital tools for creating music. So doing, it shows that Computer Music may be regarded as a major field of research and application for haptics.

## Introduction

The possible cross-empowerment of Haptics and Computer Music has not yet been widely investigated. Though, real-time digital musical instruments are a particularly promising field of research and application for the Haptics community. With an historical approach to Computer Music, this article demonstrates how Haptics correspond with a paradigm shift in our digital instruments - from the principle of *interactive musical systems* to the concept of *instrumental interaction*.

## 1 Command, Control, Mapping and Signal: Today's Concepts

The most recent approach to digital musical instruments usually exhibit three flexible components [1, 2] (fig.1):

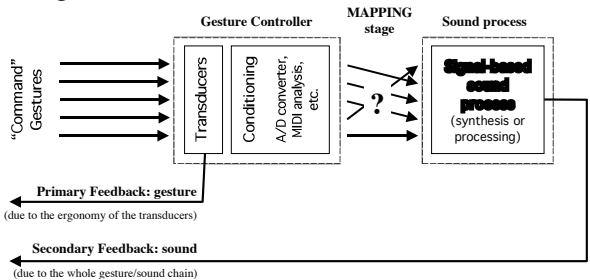

**Fig. 1:** Usual structure of a contemporary real time sound system – from [1]

1. The *gesture controller(s)*, which ergonomy impacts both the "primary feedback" and the category of the possible gestures, their dexterity, and finally expressivity.

2. The *real time sound processes*, which usually inherit from the well-known signal-based paradigms that have developed till the 90's: additive, subtractive, FM synthesis; sampling; sound filtering; sound processing; etc.
3. The *mapping stage*, which is in charge of solving the ontological gap between the gestures (or the gesture signals), and the parameters of the sound processes. The choice of an appropriate strategy is difficult, since various parameters should be varied in correlation in order to approach sufficiently thin variations in the sounds.

Such a three-level structure does extend the possibilities of our musical tools, which enabled innovative musical uses. For example, musicians can now choose the gesture controller in a large panoply (keyboards, mouth pads, joysticks, cameras, etc), and thus adapt their gesture to musical needs. Ideally, they can also program the sound quality that will be controlled or "interpreted" when performing: amplitude frequency, etc. of course, but also rhythms, timbre, localization in space, morphing…

However, we can nothing but note that the digital systems that conform with this structure *have not yet succeeded in offering expressive possibilities as interesting as those of non-digital traditional instruments* (such as the violin or the electric guitar, for example) [1]. Now that this "main stream approach" has led to a high level of complexity and technological efficiency, there must be some fundamental reasons that explain this still-remaining lack in expressivity.

## 2 A view on the traditional Instrumental relation

With traditional musical instruments (Fig. 2), apart in rare cases such as the organ pipe[1], the energy of the sounds usually comes from the player himself. The sound then results from a *gesture interaction* with an energetic coupling between the instrument and the player.

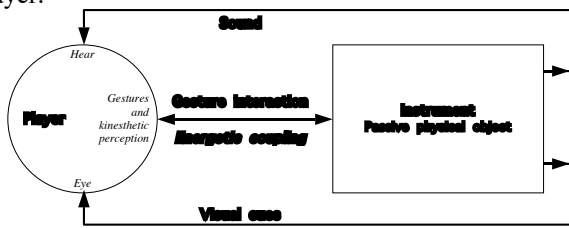

**Fig. 2:** the traditional instrumental relation

As a hypothesis, we may consider that this energetic coupling, and the tactilo-proprio-kinesthetic feedback correlated with the sound structure behavior, are important properties. They influence the sound quality and diversity, the readability of the gestures within the sound, and they are both needed for a high level of sensitivity and expressivity.

---

[1] The organ was probably the very first engine that decoupled the data flow (on/off) from the energy. Consequently, it does not fit with the analysis in the article.

The study of usual digital artifacts reinforces this hypothesis. For example, usual digital systems for sustained excitation instruments, such as strings or winds, are still hardly satisfactory for human hearing. This is not due to a lack of precision of the sound models since these are now very accurate. Indeed, with such instruments, the principles of gesture controller and mapping do encounter their limits, because they prohibit a close relationship between the player and sound, or, more precisely, between the player and the sound production mechanism[2].

## Haptics and Sound: Early Promising Results

Given the hypothesis in the above paragraph, the search for expressivity could not be solved by any improvement of any of the 3 components on Fig. 1. Haptic interfaces appear indeed to be very needed: by allowing a control of the dual force and position variables, they potentially make it possible to simulate an energetic interaction. The Fig. 3 shows how a virtual instrument may be built in order to get as close as possible to the instrumental relation of the Fig. 2.

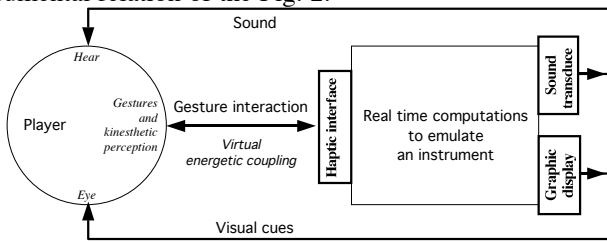

**Fig. 3:** simulation of the traditional instrumental relation

Various experiments (see [3, 4, 5, 6] for example) have nowadays proved at least partially the relevance of the structure on Fig. 3. The case study of the violin by Florens in our laboratory, that was recently improved [7], will be summarized.

In this experiment, the string was considered as a fully linear system, and the bow/string interaction implemented the most simple ever non-linear viscosity curve. Conversely to the very-simplicity in the modeling of the string, the installation implemented a TGR© haptic device [8] with a specific mechanical morphologic adapter of the ERGOS panoply [9]. As a result, most of the relevant sound cues could be easily: and naturally obtained: full excitation of the string on its first mode, full harmonic, creaking, etc. This experiment proved that *the use of a high-quality haptic system is at least as important (and probably more important) in that case than the accuracy of the computed model.*

---

[2] This analysis can be extended to the case of the piano. Whatever the accuracy of the sound signal process involved and the quality of the "touch" of the keyboard (their 'primary feedback' on Fig. 1), digital pianos are still less vivid than real ones. Indeed, they still treat the gesture as a small-band command signal for the sound process.
  The fact that the organ-like digital instruments are amongst the most comparable with their real counterparts also reinforces the hypothesis, given the foot note 1.

## Conclusion: from Control and *"Interactivity"* to *"Interaction"*

Through the discussion above, the analysis proposed in [10] that regarded Computer Graphics, can be extended to Computer Music - though with a more technical argumentation: Haptics probably play a central role in the player/sound chain, based on "preconscious capacities", whereas sound and vision are of a lesser important and "act as monitoring senses" [10].

Though, haptics interfaces should not be considered as new and eventually better gesture controlers such as those on Fig. 1. They promise a much better sensibility and expressivity of our digital instruments, hopefully comparable to those of the acoustic instruments. They correspond indeed with a *paradigm shift* in computer music. As a valid successor for the principles of Command and Control of interactive sound system, they promote the concept of *gesture Interaction* with a digital artifact through an energetically coherent bidirectional link.

A major problem, however, is then to ensure that this energetic coherence remains valid throughout the chain from gesture to sound. This calls indeed for schemes able to generate both haptic rendering and sound outputs – and eventually visual cues – in a single loop – and for further investigation.